\documentclass[journal]{IEEEtran}
\usepackage{fleqn}

\usepackage{hyperref}
\usepackage{epsfig,here}
\usepackage{graphicx}
\usepackage{cite}
\usepackage[figuresrigth]{rotating}
\begin{document}

\title{A new Slow Control and Run Initialization Byte-wise Environment (SCRIBE) for the quality control of mass-produced CMS GEM detectors\\
\vspace*{-4cm}{\tiny This work has been submitted to the IEEE Nucl. Sci. Symp. 2016 for publication in the conference record. Copyright may be transferred without notice, after which this version may no longer be available.}\vspace*{2.85cm}
}
  
\author{
S.~Colafranceschi (for the CMS muon group)

\thanks{Manuscript received November 30, 2016}

\thanks{S. Colafranceschi is with Florida Institute of Technology, Dept. of Physics and Space Sciences - 150 W. University Blvd, Melbourne, FL 32901, USA}
\thanks{* Corresponding author: scolafranceschi@fit.edu}
}

\maketitle

\begin{abstract}
The CMS collaboration aims at improving the muon trigger and tracking performance at the HL-LHC by installing new Gas Electron Multiplier (GEM) chambers in the endcaps of the CMS experiment. Construction and commissioning of GEM chambers for the first muon endcap stations is ramping up in several laboratories using common quality control protocols. The SCRIBE framework is a scalable and cross-platform web-based application for the RD51 Scalable Readout System (SRS) that controls data acquisition and analyzes data in near real time. It has been developed mainly to simplify and standardize measurements of the GEM chamber response uniformities with x-rays across all production sites. SCRIBE works with zero suppression of raw SRS pulse height data. This has increased acquisition rates to 5 kHz for a CMS GEM chamber with 3072 strips and allows strip-by-strip response comparisons with a few hours of data taking. SCRIBE also manages parallel data reconstruction to provide near real-time feedback on the chamber response to the user. Preliminary results on the response performance of the first mass-produced CMS GEM chambers commissioned with SCRIBE are presented.
\end{abstract}

\section{Introduction}
\label{sect_intro}
During 2016, the CMS\cite{:2008zzk} GEM group has started to work towards the standardization of GEM chamber\cite{Sauli:1997qp} construction across the assembly sites. The entire chamber production workload is shared among CERN and several external assembly sites. In order to meet stringent requirements on chamber performance\cite{cmsgemtdr}, standardization of the assembly lines is mandatory. After several generations of prototypes\cite{colafranceschi_ieee,colafranceschi_icatp11}, a number of quality control protocols have been defined and agreed upon in order to ensure production lines with minimal variability in manufacturing. Each assembly site is expected to run several quality control tests to validate the chamber production at all stages and to avoid construction defects or underperforming components. Among various quality controls, the chamber response test is particularly important since it characterizes chamber performance and working point. The chamber response test is also the last quality control test expected to be performed at assembly sites before shipping the chamber to CERN.

A new Slow Control and Run Initialization Byte-wise Environment (SCRIBE) has been developed to perform the chamber response quality test at all sites. It consists of measuring the analog pulse height distribution over the entire active surface of GEM chambers using the SRS\cite{Sorin}, provided by the RD51 collaboration\cite{RD51}.


\section{The framework objectives}
\label{scribe_objectives}
\begin{center}
  \begin{figure}[H]
    \resizebox{0.49\textwidth}{!}{\includegraphics{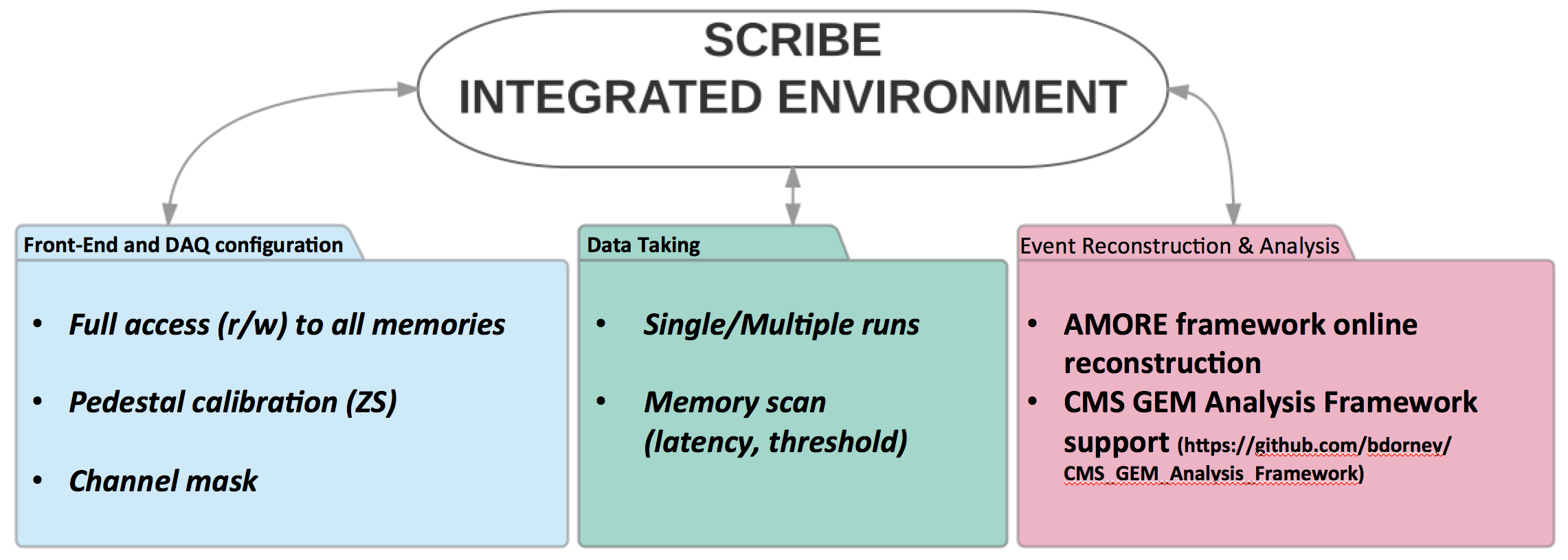}}
    \caption{The SCRIBE functionalities to enable chamber response tests at CMS GEM chamber assembly sites.}
    \label{scribeimplementation}
  \end{figure}
\end{center}

SCRIBE aims at providing a simplified and integrated (section~\ref{scribe_gui}) user experience to accomplish electronics configuration (section \ref{scribe_electronics}), data taking (section \ref{scribe_daq}), and data reconstruction (section \ref{scribe_analysis}) as depicted in Fig.~\ref{scribeimplementation}.
These three functionalities can be performed at the same time and from different devices. Since each functionality runs in a different process, SCRIBE can configure the electronics or take data while analyzing already collected data. 

Also, SCRIBE runs in a web server that accepts connections from any device through a web browser. In this way multiple users can perform multiple actions, i.e. as start new runs, analyze data or perform a read/write operation for any register of the electronics. 

Assembly sites require a near real-time data reconstruction to promptly evaluate the chambers under test and to avoid a slow down of production lines. To meet this requirement, SCRIBE has been designed with a modular and scalable architecture ({section~\ref{scribe_architecture}) with the objective to exploit parallel processing and to add an increased level of redundancy. Parallel processing is achieved by installing SCRIBE on more than one machine (SCRIBE node). Redundancy is obtained by the fact that all SCRIBE nodes are interchangeable. If one node is down or unavailable, others could still be used to perform electronics configuration and to take or analyze data. Since the chamber response quality control is expected to run at all GEM chamber assembly sites, an RPM distribution packaging system has been included to simplify SCRIBE distribution and installation.

\section{SCRIBE Architecture}
\label{scribe_architecture}
SCRIBE aims at three major goals: electronics configuration (subsection~\ref{scribe_electronics}), data taking (subsection~\ref{scribe_daq}), and data reconstruction (subsection~\ref{scribe_analysis}). The implementation of these features has been driven by framework scalability, modularity, and redundancy. 

Since the readout system (SRS) has been developed to be scalable and modular, SCRIBE has been designed to be equally scalable and modular. The SRS can handle several use-cases ranging from hundreds to tens of thousands readout channels. The Front End Concentrators (FECs), a modular building block of the SRS electronics, can work in parallel to scale up the readout system. SCRIBE can work with one or more FECs according to the use-case. The framework delivers the same performance with enhanced user experience in all scenarios. 

SCRIBE can achieve software and computer hardware redundancy by deploying the framework on several nodes. This parallel installation creates a pool of interchangeable nodes that establishes a redundant system that can run from any node. The parallel installation increases data reconstruction scalability and computing power of the reconstruction routines.

\subsection{The electronics configuration}
\label{scribe_electronics}
The readout electronics is based on the SRS with the APV25 front-end chip. The chip features 128 independent readout channels connected to the readout board of the chamber. Each channel contains a pre-amplifier and a shaper working at a frequency of 40 MHz. The analog information of the  pulses\footnote{The CMS GEM readout electronics, based on the digital VFAT chip cannot be used to record the pulses.} is multiplexed and sent to an ADC card via HDMI cables. The ADC is connected to the FEC responsible for the communication with the external devices and the control of the chips.

To configure FECs, SCRIBE needs to run on a computer that is part of the network. The configuration can be done manually (register-by-register browsing the appropriate tabs) or automatically (uploading a predefined and previously saved register snapshot). SCRIBE provides convenient read/write access to all FEC (all existing generations to date are supported) and APV register values. The electronics configuration is particularly important since the chamber quality control of GEM chambers is taking place at several research centers. To standardize the quality control outcome, all assembly sites are going to run with identical electronics settings. 

Given the required statistics to acquire highly granular chamber response data and to limit file dimension and reconstruction time, the assembly sites decided to adopt a zero-suppression (ZS) algorithm to discard data from hits with charges below a configurable threshold. ZS decreases the required bandwidth during data transfer between the FEC and the PC. This reduction is proportional to the cluster multiplicity of the average event. While a non-ZS raw data would transfer entirely a 128-value array filled with relative charges from 128 APV25 channels, ZS streams only the channels above threshold, which typically are between one and five. By zero-suppressing the data, the bandwidth used to stream raw data from the FEC to the DAQ computer is reduced by a factor $\approx 100$.

\begin{center}
  \begin{figure}[H]
    \resizebox{0.49\textwidth}{!}{\includegraphics{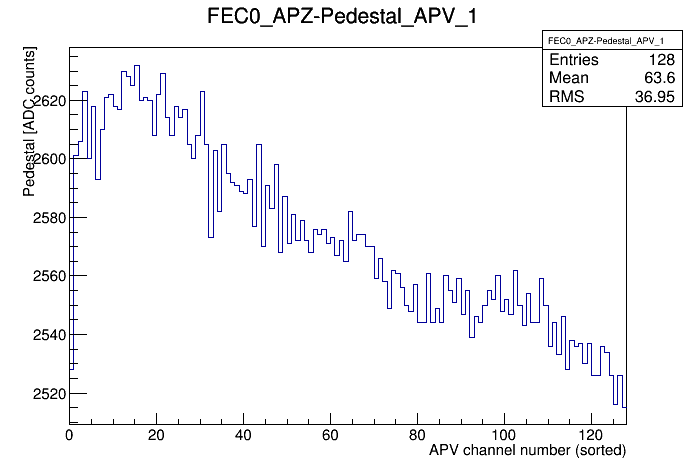}}
    \caption{A representative histogram of measured pedestals for 128 channels of one front-end APV25 chip.}
    \label{apv_pedestal}
  \end{figure}
\end{center}

The firmware of each FEC holds pedestal values (mean, Fig.~\ref{apv_pedestal}, and sigma) for each channel in order to perform pedestal subtraction from the recorded pulse heights event-by-event. The subtraction is programmed as a multiple of the sigma from the pedestal mean. 
Only data above threshold are transmitted and recorded by the DAQ. 

Pedestal values are saved in the firmware and flushed during a power cycle. SCRIBE can determine and upload these pedestal values during the initialization phase of the electronics. The user performs ZS configuration via a dedicated button that bypasses the ZS algorithm in order to actually record data on all strips to evaluate the pedestal mean and sigma. This bypass is performed channel-by-channel, chip-by-chip, and FEC-by-FEC exploiting a firmware functionality. SCRIBE, is also able to identify at runtime the number of connected chips, so that the configuration button will perform the pedestal calibration of all connected chips.

\subsection{The data acquisition system}
\label{scribe_daq}
SCRIBE has been developed to work as back-end DAQ with DATE\cite{refdate} as front-end DAQ. SCRIBE interacts with DATE via the ECS\footnote{Experiment Control System: \url{https://alice-ecs.web.cern.ch/alice-ecs/}} package. Independently of the number of SCRIBE nodes, DATE runs only in one machine (while the reconstruction runs parallel in all SCRIBE nodes, see section~\ref{scribe_analysis}). However, the availability of several DATE/SCRIBE nodes increases the system redundancy since every node can be operated as the DAQ node. Using ZS, the DAQ benefits from a variable event data stream bitrate proportional to the chamber occupancy. Without ZS a constant payload (tab.~\ref{tab:daqperf}) is sent by the FEC to the DAQ that is receiving all 128 APV25 channels (3072 considering 24 APV25 chips per GEM chamber). With ZS the FEC firmware is discarding from the payload all the channels below the defined pedestal threshold. This makes possible to record datasets with 10~M events that occupy a reasonable disk space. However, SCRIBE can seamlessly work with or without ZS since the DAQ supports both types of data payload.
\begin{table}[h]\footnotesize
	\caption{DAQ Performance (24 APV25 front-end chips, 27~APV25~time~bins, 10~M events)}
\label{tab:daqperf}
\begin{center}
  \begin{tabular}{ | c | c | c | c | c |}
    \hline
                    & Channels &                        & Theoretical  &  \\
                    & read out   &      Data                  & maximum    & Raw data \\
                    & per chamber &  streams  & event rate & file length \\
                    &  and event  & (kBytes/event) &   (kHz)         & (GBytes) \\ \hline
    Standard & 3072 & 100 & 0.12 & 960 \\ \hline
    ZS & $\approx{1-5}$ & 0.64 - 0.86 & 18 & 4 \\ \hline
  \end{tabular}
\end{center}
\end{table}


\subsection{The data reconstruction}
\label{scribe_analysis}
The data reconstruction is handled by SCRIBE through a dedicated pool where jobs are automatically submitted by the DAQ or manually submitted by users. The pool is a cloud-space where all SCRIBE nodes commit their available resources in order to parallelize data reconstruction. If SCRIBE is installed in a stand-alone machine the reconstruction is limited by the computing power of that machine, whereas deploying SCRIBE to several nodes increases the computing power of the system.

By design, any new run taken by the DAQ gets automatically submitted to the pool to be analyzed. In the pool, the first available core of the first available computer node starts to process the raw data. The pool is scalable at runtime by starting SCRIBE on other machines.

One of the most important SCRIBE features is the support of ZS data reconstruction. The data reconstruction is performed by using the AMORE framework\cite{generalamore}, in particular the SRS AMORE} developed within the RD51 collaboration. 

\begin{table}[h]\footnotesize
	\caption{AMORE reconstruction performance (24 APV25 front-end chips, 27 APV25 time bins)}
\label{tab:amorereco}
\begin{center}
  \begin{tabular}{ | c | c | c | c | c|}
    \hline
                    &  & \multicolumn{2}{|c|}{Event processing rate}\\
                    & Channels read out  & \multicolumn{2}{|c|}{(Intel Haswell 2.8GHz)}\\
                    & per chamber and event & single core & 8 cores \\ \hline
    Standard & 3072 & 50 Hz & 0.25 kHz \\ \hline
    ZS & 1-5 & 750 Hz & 3.8 kHz\\ \hline
  \end{tabular}
\end{center}
\end{table}

\begin{center}
  \begin{figure}[H]
    \resizebox{0.46\textwidth}{!}{\includegraphics{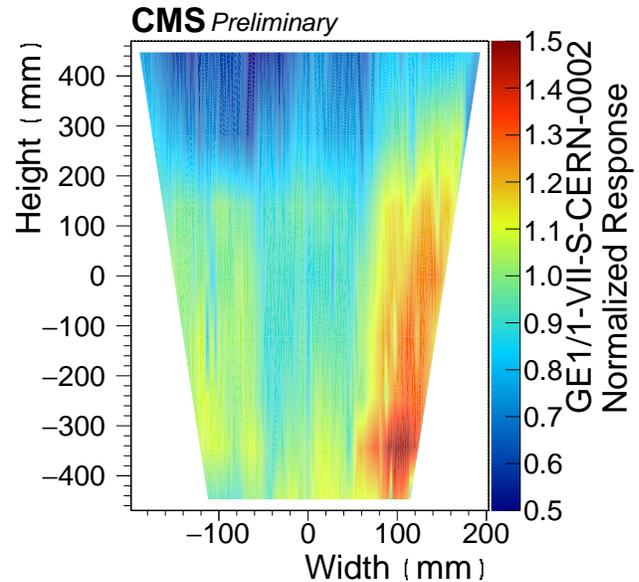}}
    \caption{A prototype chamber response, as given by the relative pulse height distribution, over the entire active surface of a CMS GEM slice test chamber, obtained using SCRIBE.}
    \label{gem_uniformity}
 \end{figure}
\end{center}

SCRIBE introduces the support for ZS into the SRS AMORE  in order to reduce the raw data file dimension (tab.~\ref{tab:daqperf}) and increase the processing rate for the reconstruction event (tab.~\ref{tab:amorereco}). Given the fact that the chamber response test aims at evaluating the performance with high granularity, high statistics runs are mandatory in order to acquire sufficient data for each of the 3072 strips. In addition to the reduced raw data file dimensions, reconstruction time is also affected by ZS. Since both disk read/write speed and RAM memory are limited, a smaller event payload can be analyzed faster. 
Once all the events are properly reconstructed, the main analysis routine (CMS GEM Analysis framework\footnote{\url{https://github.com/bdorney/CMS_GEM_Analysis_Framework}}) can run to deliver the chamber response result. Fig.~\ref{gem_uniformity} shows a typical x-y color map of pulse height information obtained using SCRIBE and the analysis routine. 

The chamber response is obtained by irradiating the chamber with x-rays\footnote{AMPTEK miniX: \url{www.amptek.com}} (silver target, 40kV, 100$\mu$A) at a gas gain of $\approx 600$ using Ar/CO$_2$ (70:30). These standard conditions are applied at all assembly sites in order to standardize all chamber response tests. Mass produced chambers are accepted if their response is within $\pm15\%$\cite{cmsgemtdr} and efficiency is above 97\%.

\section{The framework functionalities}
\label{scribe_gui}
The SCRIBE front-end has been developed using the latest dynamic web technology while the back-end relies on C/C++. Through this approach the front-end is cross-platform and cross-device, while only the back-end needs to fulfil operating system and environment requirements.
The graphical user interface provides a number of web-tabs to monitor and control the electronics, the DAQ, and the data reconstruction. 

The ``SRS general" tab supervises the most important settings of the electronics such as the SRS FEC IP addresses and ports (to be matched with the ports defined by the SRS firmware). In addition, a button allows the user to get a full snapshot of all registers associated with any SRS FECs. Such a snapshot can be used at any time to flash back the registers and restore a particular state. This is also used by SCRIBE to initialize itself and to populate all tables of all registers of all FECs/APV25s that are going to be discussed later in this section. The tabs: ``SRS system," ``ADC card," ``APV application,"  ``APV hybrid," and ``APZ registers" provide access to memories and built-in functionalities of the SRS firmware\cite{srsshort1,srsshort2}. As both registers and functionalities are FEC and APV25 front-end chip dependent, these tabs feature an additional menu to select which FEC and APV25 to communicate with. While ``SRS system" and ``ADC card" tabs provide access to memories that, in normal conditions, are commonly not frequently accessed, the ``APV application" ``APV hybrid", and ``APZ registers" tabs are designed to fully configure the front-end chip and to perform functionalities such as pedestal determination and pulse charge injection. These tabs contain tables filled with: register addresses, read-back values, values to be written, a button to execute the write, and an explicative online help. While the ``APV application" tab contains all APV25 front-end chip settings, the ``APV hybrid" tab contains the settings of the hybrid that the chip is mounted on. The ``APZ registers" tab contains a table with pedestals (mean and sigma) of the 128 channels of each APV25. These values are stored in a volatile memory of the firmware (after SCRIBE initialization) and are used during data taking to properly subtract pedestals in every event. The ``ZS pedestals" tab monitors and controls the pedestals stored in the firmware. This tab provides a channel-by-channel monitoring of all pedestal values along with summary histograms. The pedestal of each channel can be accessed and changed at runtime. This functionality can be used to mask any channel in case of need. The ``DAQ" and ``Data Reconstruction" tabs provide access to the data acquisition system and reconstruction tools. Both tabs allow to flag the SCRIBE node as DAQ node and/or as reconstruction node. While there can be only one node that performs the data taking, there is no limit to the number of nodes committed to perform the data reconstruction. The ``DAQ" tab allows to con
 the data taking and to program a set of runs to be taken. The configuration consists of declaring a number of variables such as the path where data files should be saved, chamber variables (type, id and actual HV bias current), trigger condition, run type, assembly site, run statistics and x-ray settings.
There are two ways to take data: single run or multiple runs. While in single-run mode, SCRIBE only collects the data once, during multiple-runs mode the program splits the data taking into several smaller runs and to perform a memory register scan in case of needs. A memory register scan changes the value of a control register in a predefined way each time a new run starts. The use case for splitting the measurement campaign into several smaller runs is represented by the near real-time prompt reconstruction feedback. Since raw data are taken sequentially, the reconstruction routine can start processing the already collected data while the DAQ is recording new datasets. At the end of the multiple-run data taking, the reconstructed outputs are automatically merged to provide a single deliverable. The multiple-run data taking is also useful to perform scans of any register, e.g. latency or pedestal threshold. In this case, the reconstruction routine will not merge the raw data outputs since each dataset is taken at different conditions. The ``Reconstruction" tab is designed to configure the data-processing and to provide an online reconstruction monitor of running jobs, along with a job submission system. Configuring the reconstruction consists of declaring the raw data path, the amount of statistics to process, and the quantity of cores per computer to commit to the reconstruction. Having SCRIBE deployed on several computer nodes extends the number of cores and exploits parallel processing in a very effective and scalable manner. The tab allows to enable/disable a specific computer node to be part of the reconstruction pool. In addition, from the ``Reconstruction" tab, any run can be analyzed by manually selecting the filename to be processed. A real time online feedback shows the status of the ongoing reconstructions and the jobs status. A particular job, or the entire job pool, can also be terminated via dedicated buttons. All runs present in the pool, waiting to be analyzed or being processed, are displayed and dynamically updated.

\section{Summary and conclusions}
\label{scribe_conclusions}

SCRIBE has been specifically developed for performing the chamber response test at the CMS GEM assembly sites. However, it does provide an integrated environment for any SRS based experiment. The electronics configuration, data taking and data reconstruction are integrated, monitored, and controlled online. Particular important is the ZS feature that, for large datasets ($\approx$ 10~M events), reduces data taking and reconstruction time from several weeks to a couple of hours per chamber. Without such functionality, the CMS GEM assembly sites would not be able to complete the mass production in time for the installation in the second long LHC shutdown.

SCRIBE features a web-based interface to provide ease of use and modularity to users running simple or complex setups. This enhanced user-experience simplifies the learning curve and allows new users to run the SRS from initialization to data reconstruction in a few minutes.


%
\section{Acknowledgements}
We warmly acknowledge the RD51 collaboration, the ALICE experiment and K.~Gnanvo (U.~Virginia) for the helpful collaboration. 

\end{document}